\documentclass{elsart}
\usepackage{amssymb}
\usepackage{graphicx}

\usepackage[ps2pdf,colorlinks=true]{hyperref}
 \def\ket{\!>\,} \def\ack{\,|\,}

\providecommand{\boldsymbol}[1]{\mbox{\boldmath $#1$}}
\begin{document}
\begin{frontmatter}

\title{Unified description of rotational-, $\gamma$-, and quasiparticle-band
structures in neutron-rich mass $\sim$ 110 region}

\author{ G. H. Bhat$^{1}$, J. A. Sheikh$^{1}$, Y. Sun$^{2,3,4}$, and R. Palit$^{5}$ }
\address{$^1$ Department of Physics, University of Kashmir, Srinagar,
190 006, India \\
$^2$ Department of Physics and Astronomy, Shanghai Jiao Tong
University, Shanghai 200240, People's Republic of China \\
$^3$ Collaborative Innovation Center of IFSA (CICIFSA), Shanghai
Jiao Tong University, Shanghai 200240, People's Republic of China \\
$^4$ State Key Laboratory of Theoretical Physics, Institute of
Theoretical Physics, Chinese Academy of Sciences, Beijing 100190,
People's Republic of China \\
$^5$ Department of Nuclear and Atomic Physics, Tata Institute of
Fundamental Research, Colaba, Mumbai, 400 005, India
 }

\begin{abstract}
Band structures of the neutron-rich Mo- and Ru-isotopes around A
$\sim $ 110 are investigated using the triaxial projected shell
model (TPSM) approach employing multi-quasiparticle configuration
space. The mass region under investigation depicts a rich variety of
band structures with well developed $\gamma$- and
$\gamma\gamma$-bands, and quasiparticle excitations based on them.
It is demonstrated that TPSM provides a reasonable description of
most of the observed properties, in particular, detailed structure
variations observed in Mo-isotopes are well reproduced in the
present work.
\end{abstract}
\end{frontmatter}
%\pacs{21.60.Cs, 21.10.Hw, 21.10.Ky, 27.50.+e}

%\\maketitle

\section{Introduction}
The basic modes of nuclear excitations are of rotational,
vibrational, and multi-quasiparticle (qp) in character and in many
regions of nuclear periodic table these three modes coexist. One of
the major challenges to nuclear models is to provide a unified
description of these basic excitation modes. There is a long history
of phenomenological models that were introduced to study the
interplay among the three modes of excitations \cite{BM75,VS81,AB82,BJ91}. On the microscopic front, however, there are very
few models capable of describing the three modes of excitations in a
unified manner, in particular, at higher angular momenta.

Nuclei around A $\sim$ 110 exhibit some of the most interesting
features in the nuclear periodic table. For instance, many nuclei in
this region depict quite large deformation with $\beta \sim 0.45$,
which is understood as due to the reinforcing effect of proton and
neutron deformed shell gaps at $Z=38$, 40 and $N=60$, 62
\cite{JH95,Zh07,WU04,SZ05,SZ11}. Further, in some nuclei in this
region, well developed $\gamma$- and $\gamma\gamma$- bands have been
observed up to quite high angular momenta. For example, $\gamma$-
and $\gamma\gamma$- bands have been identified in $^{104-106}$Mo
isotopes \cite{Gu95,AG96,LY01,HY04}. Although yrast bands in this
mass region have been studied using theoretical approach of total
routhian surface model
\cite{JH95,WU04,SZ05,MA91,SZ92,PM95,JS97,GA99}, there appears no
systematic investigation of the $\gamma$-bands.

Microscopic triaxial projected shell model (TPSM) approach
\cite{JS99} has been developed and it has been demonstrated to
provide a coherent and accurate description of yrast-, $\gamma$- and
$\gamma\gamma$-bands in deformed nuclei \cite{YK00}. In this
approach, the three dimensional angular-momentum projection operator
is employed to project out the good angular-momentum states from
triaxially-deformed Nilsson + BCS basis. The shell model Hamiltonian
is then diagonalized using these angular-momentum projected basis
states. This model was initially restricted to perform the
projection from the vacuum configuration only and it was possible to
study only low-spin states \cite{YK00,JY01,YJ02,PB02,SL08}. The
model space has now been expanded by including multi-qp states above
the triaxially-deformed qp vacuum and it is now possible to extend
the study of yrast and other band structures to high-spin states
\cite{GH08,JG09}.

The multi-qp TPSM approach has been employed to investigate the
high-spin band structures in Er-isotopes and in the mass $A \sim$
130 region. It has been shown in some Er-isotopes that
$\gamma$-bands built on the two-qp configurations can become yrast
at higher angular-momenta \cite{GH08}. In the mass $A \sim$ 130
region, the TPSM model has provided an alternative explanation of a
long-standing puzzle of two-aligned bands with identical intrinsic
configuration observed in some nuclei. It has been shown that the
two observed aligned states in $^{134}$Ce, having identical neutron
configuration, are the normal two-qp neutron aligned band and the
$\gamma$-band built on this configuration \cite{JG09}.

Recently, we have further generalized the TPSM approach to study the
$\gamma$-vibration in odd-proton nuclei and a preliminary
application of this new development for $^{103}$Nb has already been
reported \cite{GJ10}. In the present work, we have also developed
TPSM model for odd-neutron systems and would like to investigate the
observed $\gamma$-vibrational band structures in the mass $A \sim$
110 region where rich band structures are observed in odd-neutron
systems. This is one of the few mass regions in the periodic table
where ground-state axial-shape-asymmetry is suggested \cite{Mo06},
and also where $\gamma$- and $\gamma\gamma$- bands have been
identified in odd-mass $^{105}$Mo nucleus up to quite high angular
momenta \cite{MA91,HY04,HC02,HB06}. The purpose of the present work
is to perform a systematic study of the band structures observed in
these nuclei and, more importantly, to evaluate the transition
probabilities using the extended version of the TPSM approach.

The manuscript is organized as follows : In the next section, we
shall provide a brief sketch of the multi-qp TPSM approach, in
particular, we shall provide a few details on the multi-qp basis
states employed in the present analysis. In section III, the results
of our calculations for even-even and odd-neutron systems are
presented and discussed in two subsections. Finally, the present
work is summarized in section IV.

\section{Triaxial Projected Shell Model Approach}

In the present work, the TPSM calculations are performed for
even-even and odd-neutron isotopes. For the even-even system, the
TPSM basis is composed of 0-qp vacuum, two-proton, two-neutron and
four-qp configurations, i.e.,
\begin{eqnarray}\label{basis}
\hat P^I_{MK}\ack\Phi\ket;\\\nonumber
~~\hat P^I_{MK}~a^\dagger_{p_1} a^\dagger_{p_2} \ack\Phi\ket;\\\nonumber
~~\hat P^I_{MK}~a^\dagger_{n_1} a^\dagger_{n_2} \ack\Phi\ket;\\\nonumber
~~\hat P^I_{MK}~a^\dagger_{p_1} a^\dagger_{p_2}
a^\dagger_{n_1} a^\dagger_{n_2} \ack\Phi\ket.
\end{eqnarray}
For the study of odd-neutron system, our model space is spanned by
the following angular-momentum-projected one- and three-qp
basis$~:~$
\begin{eqnarray}\label{basisn}
\hat P^I_{MK}~a^\dagger_{n} \ack\Phi\ket ; \\\nonumber
\hat P^I_{MK}~a^\dagger_{n} a^\dagger_{p1} a^\dagger_{p2}\ack\Phi\ket ,
\end{eqnarray}
where the three-dimensional angular-momentum projection operator is
given by \cite{ringshuck}
\begin{equation}
\hat P^I_{MK} = {2I+1 \over 8\pi^2} \int~d\Omega\,
D^{I}_{MK}(\Omega)\, \hat R(\Omega),
\label{PD}
\end{equation}
with
\begin{eqnarray}
\hat R(\Omega) = e^{-\imath \alpha \hat J_z} e^{-\imath \beta \hat J_y}
e^{-\imath \gamma \hat J_z},
\end{eqnarray}
and $\ack\Phi\ket$ represents the triaxially-deformed qp vacuum
state. It is important to note that for the case of axial symmetry,
the qp vacuum state has $K=0$ \cite{KY95}, whereas, in the present
case of triaxial deformation, the vacuum state is a superposition of
all possible $K$-values. Rotational bands with the triaxial basis
states, Eqs. (\ref{basis},\ref{basisn}), are obtained by specifying
different values for the $K$-quantum number in the angular-momentum
projector in Eq. (\ref{PD}). The allowed values of the $K$-quantum
number for a given intrinsic state are determined through the
following symmetry consideration. For $\hat S = e^{-\imath \pi \hat
J_z}$, we have
\begin{equation}
\hat P^I_{MK}\left|\Phi\right> = \hat P^I_{MK} \hat S^{\dagger} \hat S
\left|\Phi\right> = e^{\imath \pi (K-\kappa)}
\hat P^I_{MK}\left|\Phi\right>.
\label{condition}
\end{equation}
The qp basis chosen above is adequate to describe high-spin states
up to $I\sim 20~\hbar$ for even-even system,  $I\sim 35/2~\hbar$ for
odd-mass nuclei. In the present analysis we shall, therefore,
restrict our discussion to this spin regime. For the self-conjugate
vacuum or 0-qp state, $\kappa=0$ and, therefore, it follows from the
above equation that only $K=$ even values are permitted for this
state. For 2-qp states, $a^\dagger a^\dagger \left|\Phi\right>$, the
possible values for $K$-quantum number are both even and odd,
depending on the structure of the qp state. For example, for a 2-qp
state formed from the combination of the normal and the
time-reversed states $\kappa = 0$, only $K$ = even values are
permitted. For the combination of the two normal states, $\kappa=1$
and only $K=$ odd states are permitted. For one-qp state,
$\kappa=1/2$ $(-1/2)$, and the possible values of $K$ are
$1/2,5/2,9/2,\dots$ ($3/2,7/2,11/2,\dots$) that satisfy Eq.
(\ref{condition}).

As in the earlier PSM calculations, we use the pairing plus
quadrupole-quadrupole Hamiltonian \cite{KY95}
\begin{equation}
\hat H = \hat H_0 - {1 \over 2} \chi \sum_\mu \hat Q^\dagger_\mu
\hat Q^{}_\mu - G_M \hat P^\dagger \hat P - G_Q \sum_\mu \hat
P^\dagger_\mu\hat P^{}_\mu .
\label{hamham}
\end{equation}
The corresponding Nilsson Hamiltonian, which is used to generate the
triaxially-deformed mean-field basis and can be obtained by using
the Hartree-Fock-Bogoliubov (HFB) approximation, is given by
\begin{equation}
\hat H_N = \hat H_0 - {2 \over 3}\hbar\omega\left\{\epsilon~\hat Q_0
+\epsilon'~{{\hat Q_{+2}+\hat Q_{-2}}\over\sqrt{2}}\right\} .
\label{nilsson}
\end{equation}
Here $\hat H_0$ is the spherical single-particle Hamiltonian, which
contains a proper spin-orbit force \cite{Ni69}. The interaction
strengths in (\ref{hamham}) are taken as follows: The QQ-force
strength $\chi$ is adjusted such that the physical quadrupole
deformation is obtained as a result of the self-consistent
mean-field HFB calculation \cite{KY95}. The monopole pairing
strength $G_M$ is of the standard form
\begin{eqnarray}\label{pairing}
G^n_M &=& {{G_1 - G_2{{N-Z}\over A}}\over A} ~{\rm for~neutrons,}\\\nonumber
G^p_M &=& {G_1 \over A} ~{\rm for~protons.}
\end{eqnarray}
In the present calculation, we take $G_1=16.22$ and $G_2=22.68$,
which approximately reproduce the observed odd-even mass difference
in the studied mass region. This choice of $G_M$ is appropriate for
the single-particle space employed in the model, where three major
shells are used for each type of nucleons: i.e. $N=3,4,5$ for
neutrons and $=2,3,4$ for protons. The quadrupole pairing strength
$G_Q$ is assumed to be proportional to $G_M$, and the
proportionality constant being fixed as 0.18. These interaction
strengths are consistent with those used earlier for the same mass
region \cite{sjb12}.

\section{Results and Discussions}

The TPSM calculations have been performed for neutron-rich even-even
and odd-neutron Mo- and Ru-isotopes which depict $\gamma$-band
structures. The calculations proceed in several stages. In the first
stage, the deformed basis are constructed from the solutions of the
triaxially-deformed Nilsson potential. The potential is solved for
each nucleus with the axial and triaxial deformation parameters,
$\epsilon$ and $\epsilon'$. The axial deformation parameter
$\epsilon$ is normally chosen from the measured quadrupole moment of
the system, wherever available, otherwise the tabulated values
\cite{Mnix} using the phenomenological potential models are
employed. The value of $\epsilon'$ is, preferably, chosen from the
minimum of the potential energy surface (PES) of the nucleus.
However, for some nuclei, PES depicts $\gamma$-softness, and for
these nuclei the value of $\epsilon'$ that reproduces the
$\gamma$-band head energy is adopted since it is known that this
band head energy is very sensitive to the non-axial deformation. In
the second stage, standard BCS equation is solved for the monopole
pairing interaction with the parameters mentioned earlier. This
defines the qp basis of our model. In the next step, states with
good angular-momentum are projected out from the qp states using the
explicit three-dimensional angular-momentum projection operator. In
the final stage, the projected basis is used to diagonalize the
shell model Hamiltonian consisting of pairing plus
quadrupole-quadrupole interaction terms in Eq. (\ref{hamham}). In
the following subsections, we shall separately present and discuss
the results for even-even and odd-neutron systems.

\subsection{Even-Even Systems}

%===============  table 1  ========================
\begin{table}
\caption{Axial and triaxial quadrupole deformation parameters
$\epsilon$ and $\gamma$ (defined by $\tan^{-1}\gamma=\epsilon' /
\epsilon$) employed in the TPSM calculation for even-even Mo and Ru
isotopes.}
\begin{tabular}{cccccccc }
\hline        & $^{102}$Mo &$^{104}$Mo     &  $^{106}$Mo  &$^{108}$Mo &$^{108}$Ru & $^{110}$Ru & $^{112}$Ru \\
\hline $\epsilon$ & 0.315 &  0.320 &    0.310    &  0.294   & 0.280    & 0.290     & 0.289 \\
       $\gamma$&25     & 22   &    20       &  25      & 29       & 28        & 25 \\\hline
\end{tabular}\label{tab1}
\end{table}
%====================================================

%==========================================================fig-1===============
\begin{figure}[htb]
 \centerline{\includegraphics[trim=0cm 0cm 0cm
0cm,width=0.7\textwidth,clip]{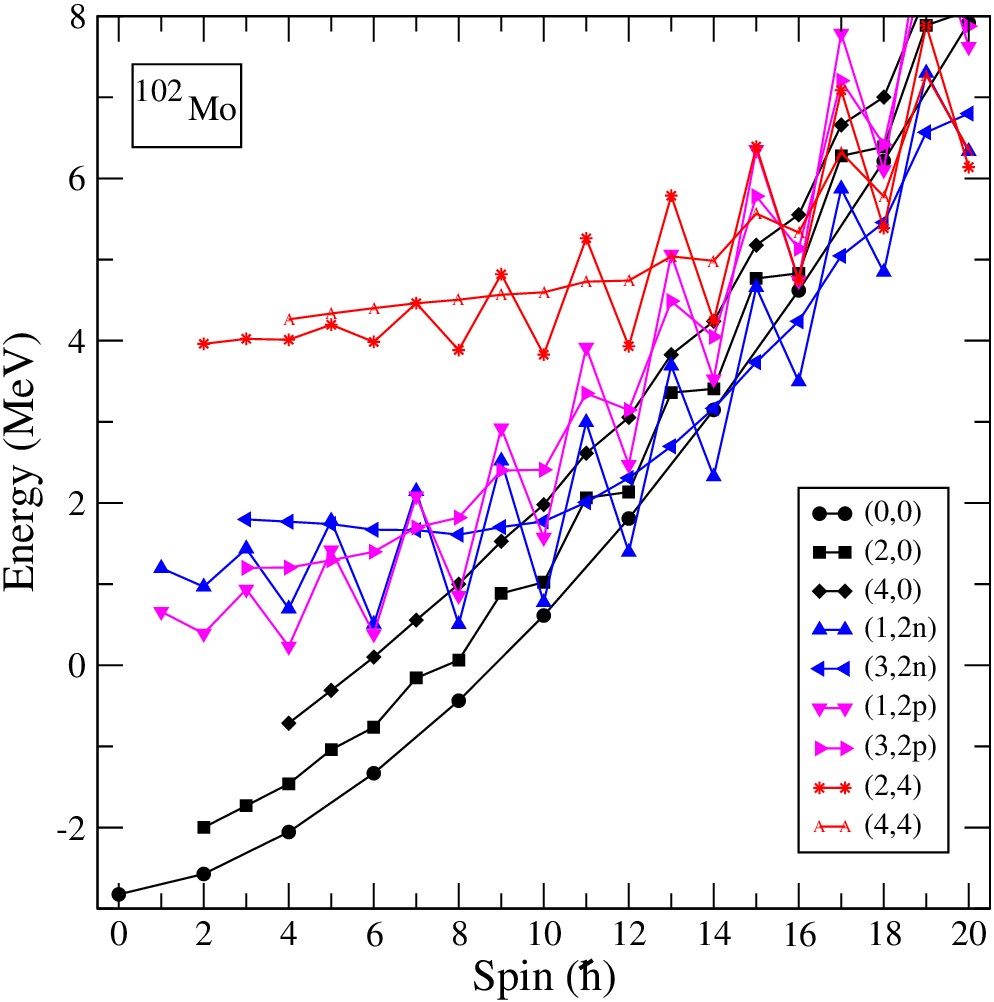}} \caption{(Color
online) Band diagram for $^{102}$Mo. The labels (0,0), (2,0), (4,0),
(1,2n), (3,2n), (1,2p), (3,2p), (2,4), and (4,4) correspond,
respectively, to the configurations of ground, $\gamma$, 2$\gamma$,
two neutron-aligned, $\gamma$-band built on this two neutron-aligned
state, two proton-aligned, $\gamma$-band built on this two
proton-aligned state, two-neutron plus two-proton aligned band, and
$\gamma$-band built on this four-qp state.} \label{fig1}
\end{figure}
%====================================================================
%===============  fig.2=====================================================
\begin{figure}[htb]
 \centerline{\includegraphics[trim=0cm 0cm 0cm
0cm,width=0.620\textwidth,clip]{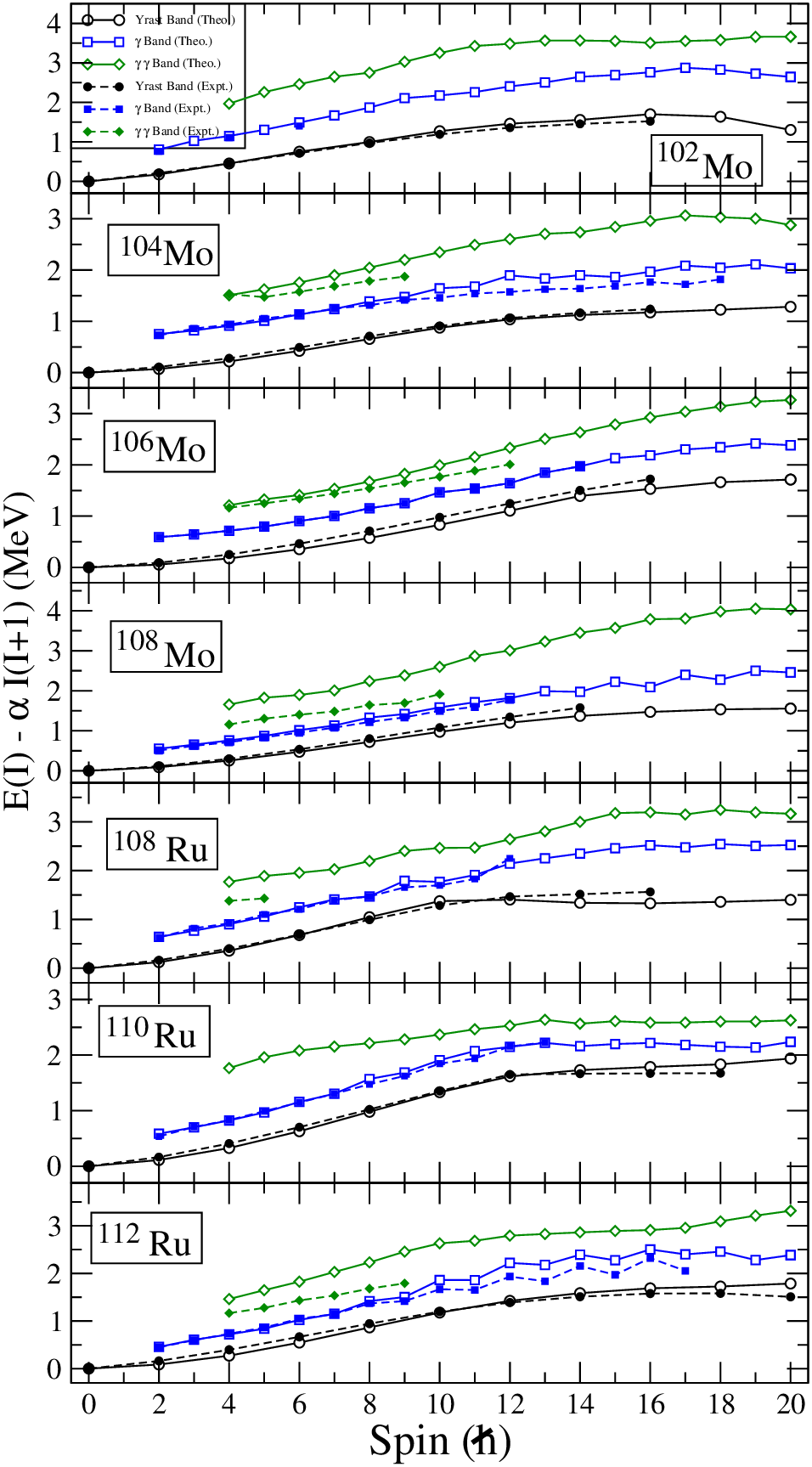}}
\caption{(Color online) Comparison of the measured energy levels of
Yrast-, $\gamma$-, and $\gamma$$\gamma$-bands for $^{102,104,108}$Mo
and $^{108,110}$Ru nuclei and the results of TPSM calculations. The
scaling factor $\alpha$ appearing in the y-axis is defined as
$\alpha=32.32 A^{-5/3}$. Data are taken from
Refs.~(\cite{AG96,HY04,LY01,CH04,JZ03}.
 }
\label{fig2}
\end{figure}
%====================================================================

For even-even systems, we have carried out TPSM calculations for
$^{102-108}$Mo and $^{108-112}$Ru isotopes. The axial and triaxial
deformation parameters $\epsilon$ and $\gamma$ used for these nuclei
are listed in Table \ref{tab1}. The axial deformations have been
chosen from the earlier studies
\cite{def1mo,def2mo,def108ru,def112ru} and triaxial deformations are
chosen, as already mentioned earlier, in such a way that the
bandhead of the $\gamma$-bands is reproduced.

The basis space for even-even systems is composed of 0-qp,
two-quasineutron, two-quasiproton and two-quasineutron plus
two-quasiproton configurations as given in Eq. (\ref{basis}). The
projected bands from 0-qp configuration has $\kappa=0$ and the
allowed values for the projected configurations are $K =
0,2,4,\dots$. The projected band for $K=0$ constitutes the main
component of the ground-state band of an even-even system and the
projected bands from $K=2$ and $K=4$ correspond to the main
components of the $\gamma$- and $\gamma\gamma$-band, respectively.
The projected bands from two-qp states can have either $\kappa=0$ or
1. The configurations with $\kappa=0$ do not become favored for the
spin-regime studied in the present work and, therefore, have been
disregarded. However, the two-qp configurations with $\kappa=1$,
referred to as the aligned states, become yrast at higher
angular-momenta. As an illustrative example, Fig.~\ref{fig1} depicts
the projected energies before configuration mixing from different qp
configurations for $^{102}$Mo. This figure, referred to as the band
diagram \cite{KY91}, is quite instructive to unravel the intrinsic
structures of the rich band structures observed in rotating nuclei.
To simplify the discussion on such diagrams, the bands are labeled
by the symbol $(K, \# t)$, where $"K"$ is the projection of angular
momentum in the intrinsic frame and the symbol $"\#"$ denotes the
number of qps with $t=n(p)$ for neutrons (protons). The ground-state
band in Fig.~\ref{fig1}, as expected, has the configuration $(0,0)$
and is crossed by the two-neutron aligning state $(1,2n)$ at angular
momentum $I=12$. Although the two-proton aligned band, $(1,2p)$, is
almost at the same excitation energy as that of $(1,2n)$ at lower
angular momenta, it does not cross the ground-state band. Bandheads
of $\gamma$- and $\gamma\gamma$-bands are at excitation energies of
0.83 MeV and 2.11 MeV, respectively.

%%===============  fig.3 =====================================================
\begin{figure}[!htb]
 \centerline{\includegraphics[trim=0cm 0cm 0cm
0cm,width=0.7\textwidth,clip]{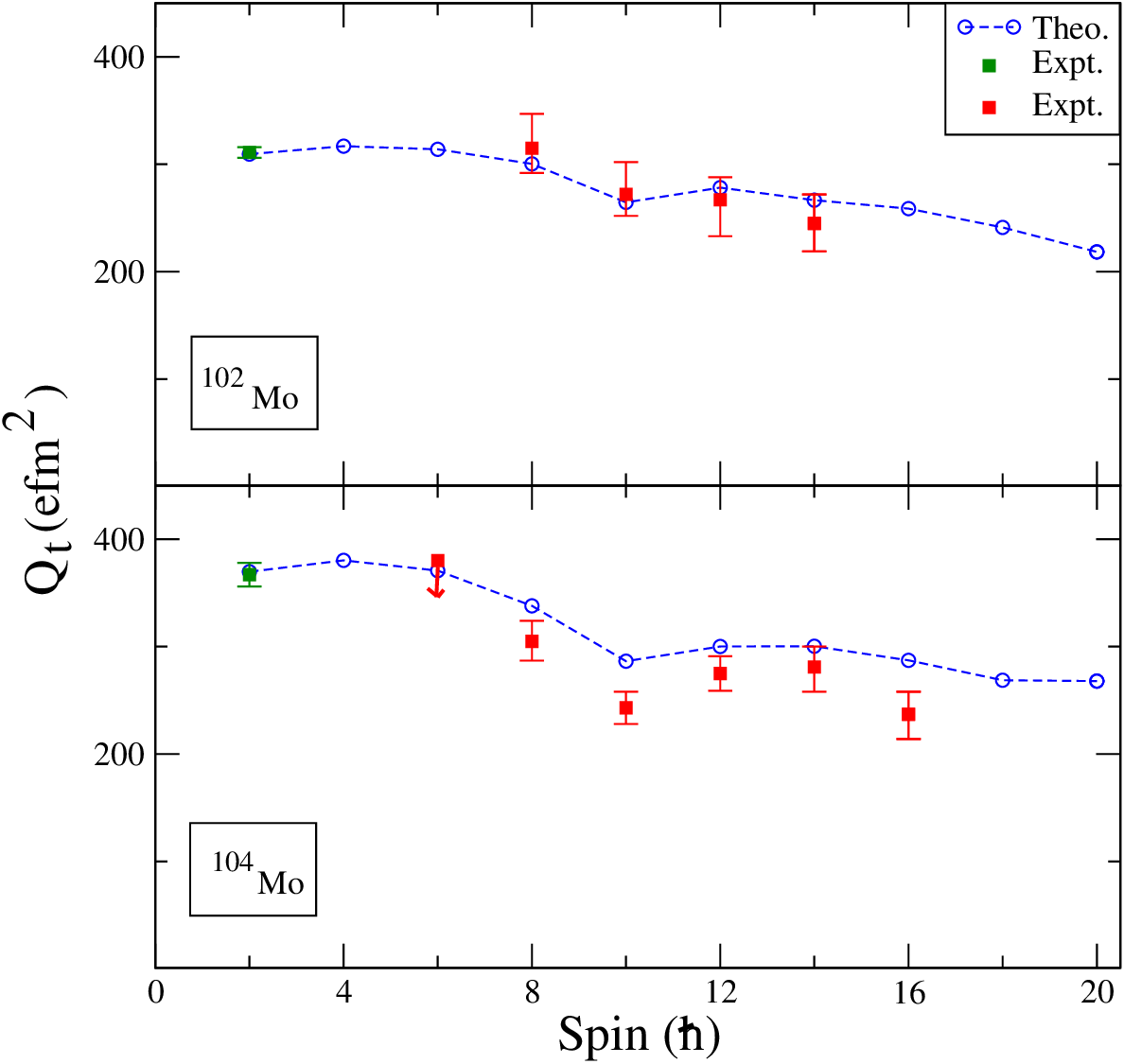}} \caption{(Color
online) Comparison of experimental and calculated $Q_t$ for $^{
102,104}$Mo. There are two sets of experimental data - one from Ref.
\cite{Raman87} (shown in green symbols) and the other from Ref.
\cite{Snyder13} (shown in red symbols). } \label{fig3}
\end{figure}
%==============================================================================
%%===============  fig.4 =====================================================
\begin{figure}[!htb]
 \centerline{\includegraphics[trim=0cm 0cm 0cm
0cm,width=0.7\textwidth,clip]{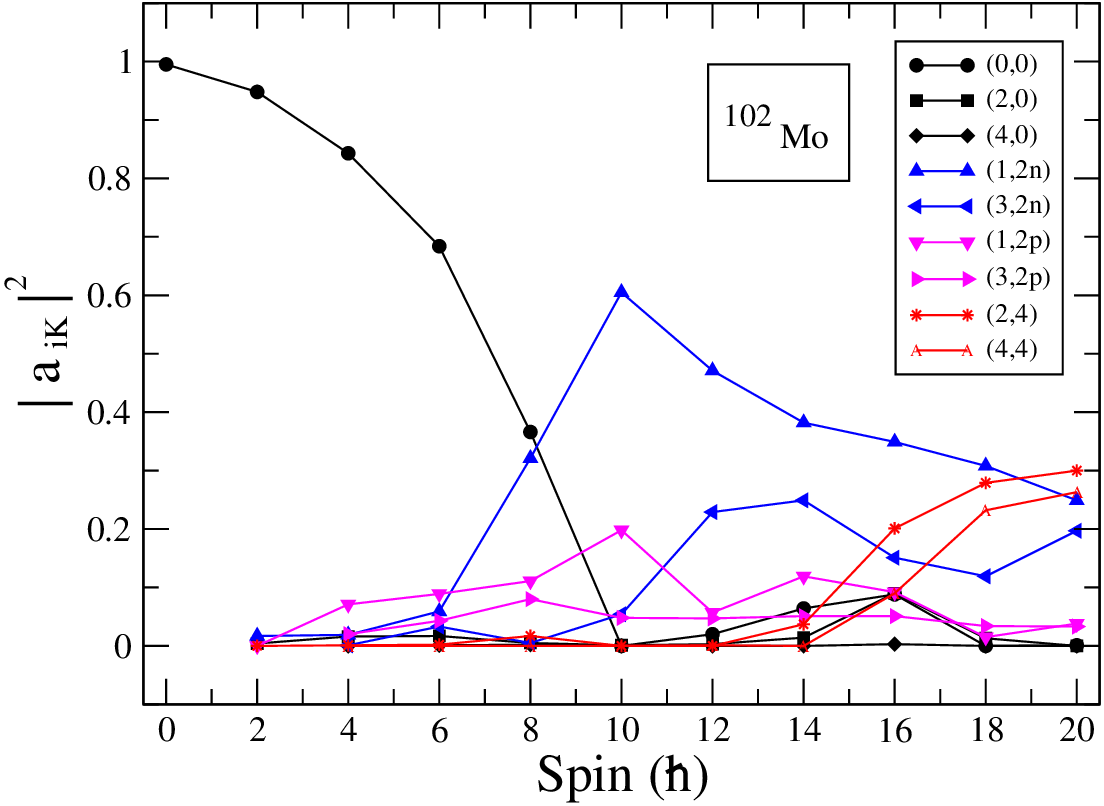}} \caption{(Color
online) Probability of various projected K-configurations in the
wave function of the Yrast band of $^{102}$Mo. The labels (0,0),
(2,0), (4,0), (1,2n), (3,2n), (1,2p), (3,2p), (2,4), and (4,4)
correspond, respectively, to the configurations of ground, $\gamma$,
2$\gamma$, two neutron-aligned, $\gamma$-band built on this two
neutron-aligned state, two proton-aligned, $\gamma$-band built on
this two proton-aligned state, two-neutron plus two-proton aligned
band, and $\gamma$ band built on this four-qp state.
  }
\label{fig4}
\end{figure}
%====================================================

The TPSM results obtained after configuration mixing are compared
with the known experimental data for the studied even-even Mo- and
Ru- isotopes in Fig.~\ref{fig2}. For $^{102}$Mo, the yrast band is
experimentally known up to angular-momentum $I=16$ and very few
states have been observed for the $\gamma$-band. It is evident from
Fig.~\ref{fig2} that TPSM reproduces the known experimental data
reasonably well, in particular, the $\gamma$-bandhead energy which
approximately lies at {\rm 1 MeV}. There is no known data for the
$\gamma\gamma$-band in this isotope and the TPSM predicts the
bandhead of this band at $\sim$ 2 MeV excitation energy. For
$^{104,106,108}$Mo, the experimental data is quite rich, with both
$\gamma$- and $\gamma\gamma$- bands known up to quite high
angular-momenta. It is quite evident from the comparison that TPSM
reproduces the yrast-, $\gamma$-, and $\gamma\gamma$- bands
surprisingly well. For $^{108}$Mo, there is a clear discrepancy
between the theory and the experimental energies of the
$\gamma\gamma$-band. This problem has also been noted in the TPSM
study of some other isotopes \cite{GJ10}.

From the TPSM results of Ru-isotopes with $A=108$ to 112, it is
again evident from Fig.~\ref{fig2} that calculations reproduce the
known experimental data quite well, especially for the known yrast-
and $\gamma$-bands. $\gamma$-band. In $^{108}$Ru these are known up
to spin $I=12$ and, on the other hand, only two states with $I=4$
and 5 have been identified in the $\gamma\gamma$-band. For
$^{110}$Ru, $\gamma$-band is known up to $I=13$ and no states in
$\gamma\gamma$-band have been identified. Both $\gamma$- and
$\gamma\gamma$-bands are known up to quite high spin in $^{112}$Ru.
The discrepancy in the bandhead energy of the $\gamma\gamma$-band is
again noted for Ru-isotopes.

%====================================================
% \begin{widetext}
\begin{table}
\caption{Comparison of the known experimental $\gamma$-band $Q_t$
values (in efm$^2$) and calculated ones for $^{102,104,108}$Mo and
$^{108,110}$Ru isotopes.}
%\scriptsize{
\begin{tabular}{cccccc}
\hline $(I,K)_i\rightarrow (I,K)_f$ & $^{102}$Mo  & $^{104}$Mo   &
$^{106}$Mo  & $^{108}$Ru & $^{110}$Ru
 \\
%         & Expt. & Theo. &Expt. & Theo.&Expt.&Theo.&Expt.& Theo.\\
\hline
  %******************yrast---->yrast*************************************************************
$(4,2)_i\rightarrow (2,2)_f$ & 188.82    & 237.56                                        &  151.38      & 197.27    & 210.06     \\
$(5,2)_i\rightarrow (3,2)_f$ &243.20     & 264.55                                        &  288.03     & 259.40    &275.46     \\
$(6,2)_i\rightarrow (4,2)_f$ & 236.62    & 312.84                                          &  296.15      &253.71     &273.54     \\
 Expt.&     &                                      &        &    &  278$^{+52}_{-37}$   \\

$(7,2)_i\rightarrow (5,2)_f$ & 270.50    &298.74                                           &  229.99     &296.68     &305.57    \\
Expt. &     &                                         &       &    &  292$^{+37}_{-31}$ \\

$(8,2)_i\rightarrow (6,2)_f$ &245.36     & 329.45       &  209.13     & 312.42      &283.06   \\
Expt. &    &    300$^{+100}_{-80}$&           &    330$^{+90}_{-70}$   & 268$^{+40}_{-31}$   \\

$(9,2)_i\rightarrow (7,2)_f$ & 154.19    & 266.53           & 297.07  &309.98    & 310.51    \\
Expt. &     &    233$^{+59}_{-34}$ &        310$^{+80}_{-80}$ &    &   299$^{+69}_{-48}$  \\

$(10,2)_i\rightarrow (8,2)_f$ & 185.66   & 330.83           & 312.88      &327.14    &278.46    \\
Expt. &    &     312$^{+49}_{-42}$ &            &  317$^{+64}_{-55}$  &  \\

$(11,2)_i\rightarrow (9,2)_f$ &255.92    & 282.12                           & 256.00      &305.10    &336.44    \\
Expt. &   &     267$^{+30}_{-30}$ &                      &      &    \\
$(12,2)_i\rightarrow (10,2)_f$ & 315.20  & 324.98                                         & 314.55      &326.81    &268.33     \\
$(13,2)_i\rightarrow (11,2)_f$ & 273.78  &298.49                                    & 260.70      &280.54    &339.28     \\

$(14,2)_i\rightarrow (12,2)_f$ & 254.59  & 317.68                                         & 258.33      &354.90      &90.82    \\
$(15,2)_i\rightarrow (13,2)_f$ & 282.65  &319.88                                          & 261.57      &280.03     &339.00    \\

$(16,2)_i\rightarrow (14,2)_f$ &273.83   & 115.91                                          & 268.77      &370.64     &365.87     \\
$(17,2)_i\rightarrow (15,2)_f$ & 285.32  &313.70                                         &  262.34     &264.17     &335.28     \\

$(18,2)_i\rightarrow (16,2)_f$ & 216.05  &396.70                                         &  267.72     &372.42     &369.11      \\
$(19,2)_i\rightarrow (17,2)_f$ & 276.24  & 296.71                                       & 292.36      &240.13     &327.13      \\

$(20,2)_i\rightarrow (18,2)_f$ & 321.25  &430.03                                         & 297.30      &373.71      &370.38     \\
\hline%-------------------------------------------------------------------------------------------------------------------------------
$(4,2)_i\rightarrow (2,0)_f$   & 95.088  &  64.45                          &  84.18     & 68.71  & 71.62     \\
Expt.  &   &                     &             & 78$^{+26.5}_{-26}$&    \\
$(6,2)_i\rightarrow (4,0)_f$   & 98.34   & 44.81                                        &  96.34      &52.43   &46.81     \\
$(8,2)_i\rightarrow (6,0)_f$   & 81.59  & 35.16                                           &  80.34      &40.87    &31.32      \\
$(10,2)_i\rightarrow (8,0)_f$  & 57.53  & 24.31                                         &  85.11      &50.83    &22.94    \\
$(12,2)_i\rightarrow (10,0)_f$ & 62.22  &  8.79                                      &  64.08      &78.35    &23.69     \\
$(14,2)_i\rightarrow (12,0)_f$ & 53.11  & 12.68                                       &  37.11      &89.88    &46.55    \\
$(16,2)_i\rightarrow (14,0)_f$ & 44.32  & 28.07                                           &  30.55       &30.04     &59.74    \\
$(18,2)_i\rightarrow (16,0)_f$ & 34.89  &29.59                                         &  26.95       &52.69    &61.92     \\
$(20,2)_i\rightarrow (18,0)_f$ & 30.21  & 18.86                                          &  73.35       &70.23    &44.86     \\
\hline
\end{tabular}%}
\label{tab2}
\end{table}
%\end{widetext}
%\end{tabular}
%\end{table*}
%\end{widetext}}
%===========================================================

The $Q_t$ transition probabilities have been evaluated using the
expressions already given in our earlier publications
\cite{GH14,GH12}. As examples, the calculated $Q_t$ transition
probabilities for two Mo-isotopes are depicted in Fig.~\ref{fig3}
(and for other Mo- and Ru-isotopes they have been discussed in
Ref.~\cite{Chanli15}). The transitions were evaluated using the
standard effective charges of $e_n=0.5e$ and $e_p=1.5$. The measured
value of $Q_t$ for $I=2$ in Fig.~\ref{fig3} is from Ref.
\cite{Raman87} and for other spin values is from the recent detailed
lifetime analysis \cite{Snyder13}.  The drop in the BE2 transitions
around $I=8-10$ is noted for both the nuclei in Fig.~\ref{fig3} and
these features are correctly described by the present TPSM
calculations.

In order to correlate the above observed behavior of $Q_t$ with the
structural changes along the yrast line, the projected
K-configurations in the wave function of the yrast band are plotted
in Fig.~\ref{fig4} for $^{102}$Mo as an illustrative example. It can
be seen that the yrast band up to $I=6$ is dominated by the
qp-vacuum configuration with $K=0$, and above this spin value, it is
noted that the two-quasineutron state with $K=1$ becomes dominant
and this change in the yrast band structure gives rise to the
observed drop in the measured $Q_t$. It is also evident from
Fig.~\ref{fig4} that above $I=14$, four-qp configurations having
$K=2$ and 4 become important, which leads to further drop in $Q_t$
with increasing spin. Thus the successive changes in the structure
associated with spin alignments of nucleons roughly result in a 1/3
of reduction in collectivity of the yrast band at $I=20$.

%===========================%===============  table 3  ========================
\begin{table}
\begin{center}
\caption{Axial and triaxial quadrupole deformation parameters
$\epsilon$ and $\gamma$ (defined by $\tan^{-1}\gamma=\epsilon' /
\epsilon$) employed in the TPSM calculation for odd-neutron Mo and
Ru isotopes.}
\begin{tabular}{cccccccc }
\hline        & $^{103}$Mo &$^{105}$Mo  &$^{107}$Mo &$^{109}$Ru  & $^{111}$Ru   \\
\hline $\epsilon$ & 0.345    &  0.316    &  0.300   & 0.281    & 0.263      \\
       $\gamma$&23        & 20        &  30      & 22       & 23         \\\hline
\end{tabular}\label{tab3}
\end{center}
\end{table}
%===========================%===============  table 3  ========================
%===================================fig 4=====================================%===============  ========fig. 11============================================
\begin{figure}[htb]
 \centerline{\includegraphics[trim=0cm 0cm 0cm
0cm,width=0.7\textwidth,clip]{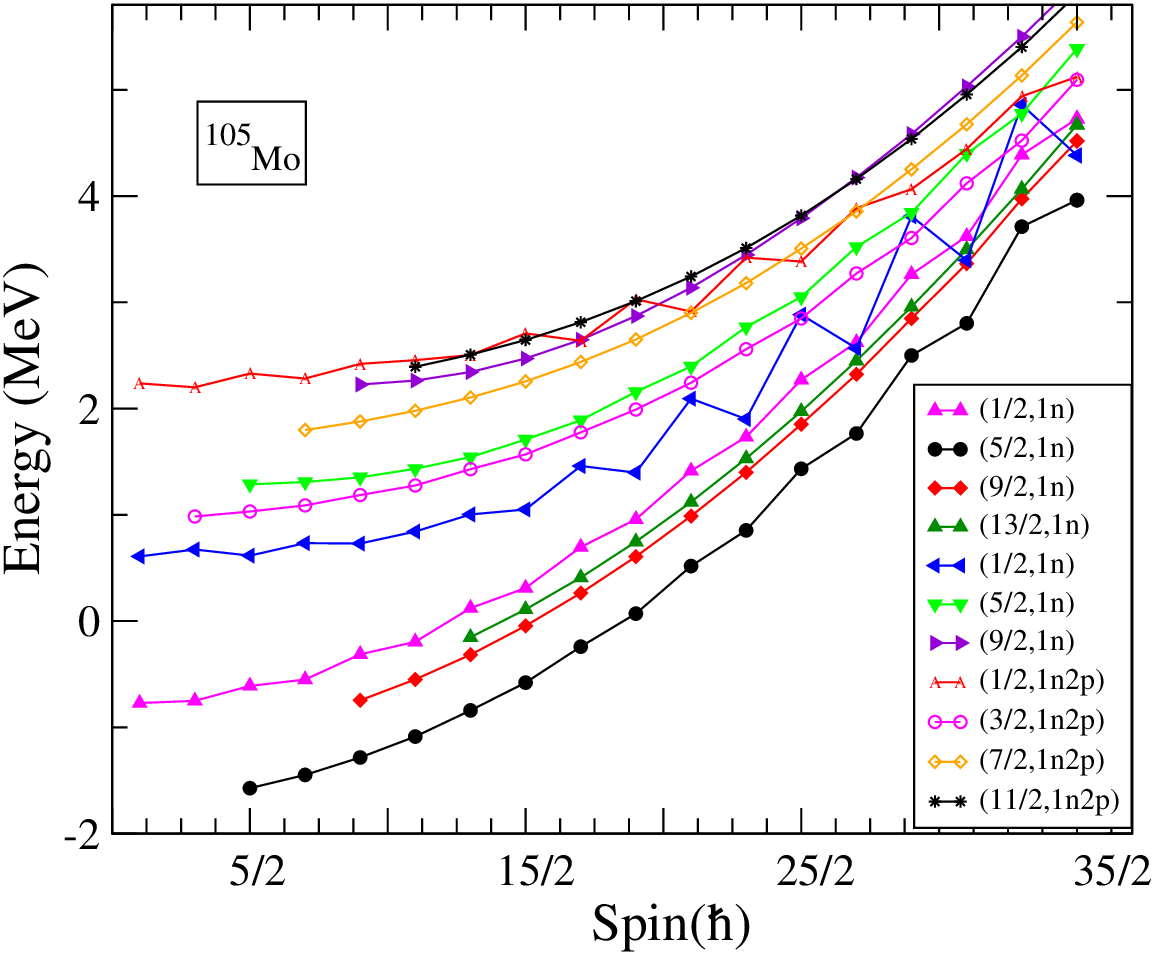}} \caption{(Color
online) Band diagram for $^{105}$Mo depicting the angular-momentum
projected bands from one- and three-qp states. For
clarity, only the lowest projected bands are shown and in the
numerical calculations, projection has been performed from
forty-four intrinsic states for this nucleus. } \label{fig5}
\end{figure}
%==============================================================================
%=================================fig-6====================================
\begin{figure}[htb]
 \centerline{\includegraphics[trim=0cm 0cm 0cm
0cm,width=0.62\textwidth,clip]{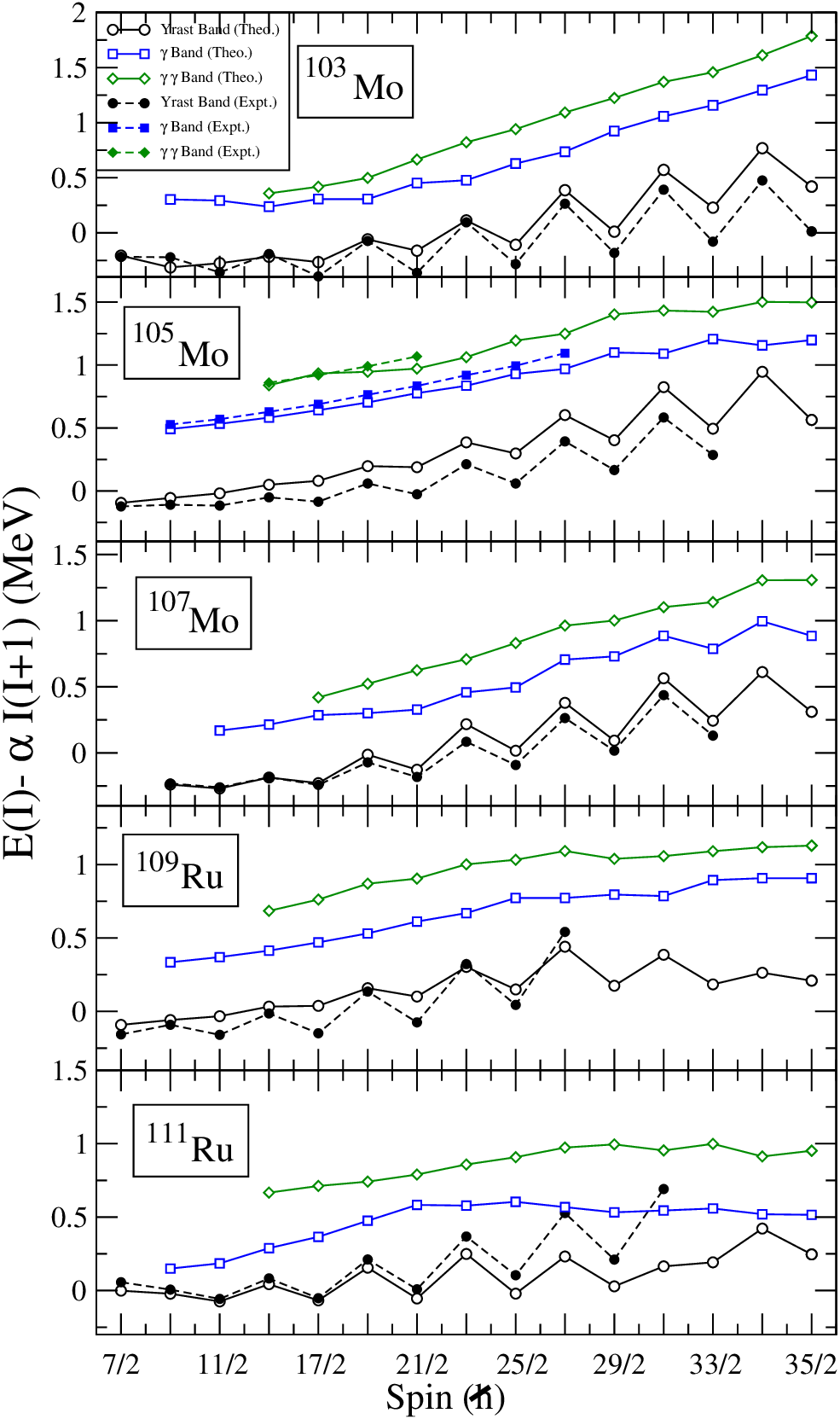}}
\caption{(Color online)  Comparison of the measured energy levels of
Yrast- $\gamma$- and $\gamma$$\gamma$-Bands for $^{103,105,107}$Mo
and $^{109,111}$Ru nuclei and the results of TPSM calculations. The
scaling factor $\alpha$ appearing in the y-axis is defined as
$\alpha=32.32 A^{-5/3}$. Data taken from Refs. \cite{HB06,JK98}.}
\label{fig6}
\end{figure}
%=========================================================================
%==============================fig 5==========================================%===============  ========fig. 11============================================
\begin{figure}[htb]
 \centerline{\includegraphics[trim=0cm 0cm 0cm
0cm,width=0.7\textwidth,clip]{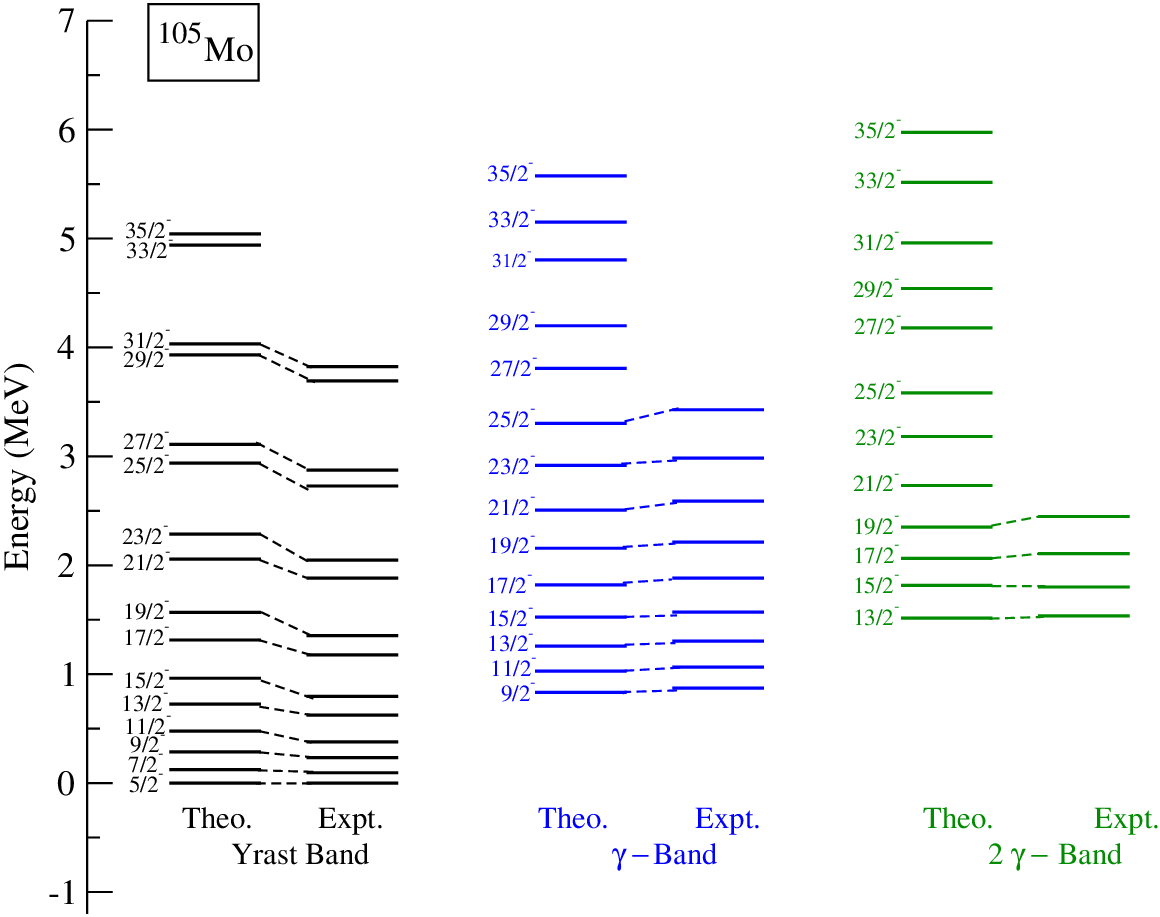}} \caption{(Color
online) The calculated yrast-, $\gamma$-, and $\gamma\gamma$-bands of
$^{105}$Mo are compared with the corresponding experimental data
\protect\cite{HB06}.} \label{fig7}
\end{figure}
%==============================================================================
%=%===============  ======fig. 7 ==============================================
%=\begin{figure}[htb]
%= \centerline{\includegraphics[trim=0cm 0cm 0cm
%=0cm,width=0.52\textwidth,clip]{qt_odn.eps}}
%=\caption{(Color online)The calculated Q$_t$s for Mo- and Ru-isotopes.} \label{fig7}
%=\end{figure}
%==============================================================================
 %\begin{widetext}
\begin{table}{\scriptsize{
\caption{Predicted B(E2) values (in W.u.) for $^{103,105,107}$Mo
and $^{109,111}$Ru isotopes.}
%\scriptsize{
\begin{tabular}{cccccc}
\hline $(I,K)_i\rightarrow (I,K)_f$ & $^{103}$Mo  & $^{105}$Mo   & $^{107}$Mo   &$^{109}$Ru  & $^{111}$Ru \\
\hline
 %******************yrast---->yrast*************************************************************
$(9/2,5/2)_i\rightarrow (5/2,5/2)_f$    &10.10  & 12.36        &               & 11.05         &  8.38     \\
$(11/2,5/2)_i\rightarrow (7/2,5/2)_f$   &40.46  & 21.24        & 20.32         &  32.41        &  23.49       \\
$(13/2,5/2)_i\rightarrow (9/2,5/2)_f$   &60.10  & 42.97        & 50.30         &  62.84        &  49.05          \\
$(15/2,5/2)_i\rightarrow (11/2,5/2)_f$  &92.07  & 63.76        & 107.3         &  108.68        & 92.75           \\
$(17/2,5/2)_i\rightarrow (13/2,5/2)_f$  &119.77 & 98.97        & 100.57         & 80.62         &  90.36         \\
$(19/2,5/2)_i\rightarrow (15/2,5/2)_f$  &135.60 & 119.39        &120.23          &109.43          & 94.88        \\
$(21/2,5/2)_i\rightarrow (17/2,5/2)_f$  &149.69 & 134.77        &145.71          &101.01          & 101.37          \\
$(23/2,5/2)_i\rightarrow (19/2,5/2)_f$  &161.45 & 146.48        &155.16          &114.71          & 104.17          \\
$(25/2,5/2)_i\rightarrow (21/2,5/2)_f$  &160.37 & 145.45        &156.67          &110.31          &100.60           \\
$(27/2,5/2)_i\rightarrow (23/2,5/2)_f$  &170.71 & 153.84        &164.55          &121.71          &110.16            \\
$(29/2,5/2)_i\rightarrow (25/2,5/2)_f$  &168.65 & 152.95        &165.36          &106.31          & 99.80           \\
$(31/2,5/2)_i\rightarrow (27/2,5/2)_f$  &175.84 & 143.54        &169.82          &108.94          &107.10            \\
$(33/2,5/2)_i\rightarrow (29/2,5/2)_f$  &174.98 &  142.72       &171.30          &98.86          & 90.63          \\
$(35/2,5/2)_i\rightarrow (31/2,5/2)_f$  &157.81 & 115.03        &147.79          &99.30          & 76.94           \\\hline
 %******************gamma---->gamma*************************************************************
$(13/2,9/2)_i\rightarrow (9/2,9/2)_f$   & 42.08 & 36.67        &                &  13.93        &   21.81           \\
$(15/2,9/2)_i\rightarrow (7/2,9/2)_f$   & 60.72 & 72.46        & 83.57         &  64.92        &   55.62            \\
$(17/2,9/2)_i\rightarrow (13/2,9/2)_f$  & 119.77&115.32         &103.42          & 101.09         & 84.57           \\
$(19/2,9/2)_i\rightarrow (15/2,9/2)_f$  &125.02 &108.72         &131.36          & 85.64         &  70.54             \\
$(21/2,9/2)_i\rightarrow (17/2,9/2)_f$  &134.69 &123.55         &111.32          & 83.11         & 75.82               \\
$(23/2,9/2)_i\rightarrow (19/2,9/2)_f$  &126.69 &106.48         &114.41          & 78.72         & 70.96               \\
$(25/2,9/2)_i\rightarrow (21/2,9/2)_f$  &147.43 &129.68         &132.83          & 98.59         & 89.73             \\
$(27/2,9/2)_i\rightarrow (23/2,9/2)_f$  &139.22 &114.49         &126.17          & 85.52         & 76.47             \\
$(29/2,9/2)_i\rightarrow (25/2,9/2)_f$  &155.79 &125.54         &146.49          & 111.11         &85.44             \\
$(31/2,9/2)_i\rightarrow (27/2,9/2)_f$  &134.61 &119.07         &127.53          & 96.26         & 97.10              \\
$(33/2,9/2)_i\rightarrow (29/2,9/2)_f$  &125.32 &112.03         &129.99          & 105.52         &80.63             \\
$(35/2,9/2)_i\rightarrow (31/2,9/2)_f$  &130.41 &106.15         & 124.30         & 86.50         & 70.79          \\  \hline
 %******************2gamma---->2gamma*************************************************************

$(17/2,13/2)_i\rightarrow (13/2,13/2)_f$  &43.97 &   42.46        &               &22.43          &  25.27          \\
$(19/2,13/2)_i\rightarrow (15/2,13/2)_f$  &80.51 &  77.26         &82.96          & 85.64         &  73.67            \\
$(21/2,13/2)_i\rightarrow (17/2,13/2)_f$  &133.60 & 125.70        &149.51          & 97.29         &  87.52            \\
$(23/2,13/2)_i\rightarrow (19/2,13/2)_f$  &126.96 & 108.74        &160.00          & 109.93         & 96.61            \\
$(25/2,13/2)_i\rightarrow (21/2,13/2)_f$  &146.53 & 138.63        &149.62          & 98.21         &  87.94           \\
$(27/2,13/2)_i\rightarrow (23/2,13/2)_f$  &136.78 & 120.82        &117.85          & 93.71         &   76.47           \\
$(29/2,13/2)_i\rightarrow (25/2,13/2)_f$  &124.02 & 140.48        &121.05          & 100.13         &  97.91            \\
$(31/2,13/2)_i\rightarrow (27/2,13/2)_f$  &147.74 & 124.89        &136.08          & 87.11         &   80.34          \\
$(33/2,13/2)_i\rightarrow (29/2,13/2)_f$  &150.57  &130.73        &147.80          & 85.52         &  87.04            \\
$(35/2,13/2)_i\rightarrow (31/2,13/2)_f$  &134.59  &91.14         &142.43          & 70.50         &  78.62            \\
\hline
\end{tabular}\label{tab4}}}
\end{table}
%\end{widetext}
%===========================================================

In Table~\ref{tab2}, we show the calculated $Q_t$ quadrupole
transition moments along the $\gamma$-band. The inter-band
transitions from the $\gamma$ to the yrast-band are also shown.
Previously inter-band transitions between the $\gamma$ and the
yrast-band were studied for some rare-earth nuclei in a smaller
configuration space without inclusion of qp excitations \cite{PB02}.
The experimental $Q_t$ values known for some transitions are also
depicted in Table~\ref{tab2}. It is noted that these $Q_t$ values
are well reproduced by the TPSM calculation. $Q_t$ transitions along
the $\gamma$-band are noted to increase with increasing spin and the
transition quadrupole moment for the odd-spin value is slightly
smaller than for the even-spin value. It is found that the
inter-band transitions are in general 3-5 times weaker than those
in-band transitions, but still they are large enough to claim a
similar collective structure of the $\gamma$-band as for the
yrast-band.

\subsection{Odd-neutron systems}

The TPSM calculations have been performed for odd-neutron
$^{103-105}$Mo and $^{109,111}$Ru isotopes by using the same
coupling parameters in the Hamiltonian (\ref{hamham}) as for the
even-even system. The configuration space consists of 1- and 3-qp
states as given in (\ref{basisn}). The axial and non-axial
deformations used in this calculation are listed in Table
\ref{tab3}. The axial deformations have been adopted from earlier
studies on these nuclei
\cite{def1mo,def2mo,def108ru,def112ru,def103mo}. The triaxial
deformation $\epsilon'$, shown in the form of $\gamma$ in Table
\ref{tab3}, are chosen in such a way that the bandhead of the
$\gamma$ bands in these odd-neutron isotopes is reproduced.

In Fig.~\ref{fig5}, we show the calculated band diagram for
$^{105}$Mo as the representative example for odd-neutron systems.
The ground-state is the projected 1-qp state with $K=5/2$. For
odd-mass nuclei, there are two possible $\gamma$-bands with
$K=K_g{\pm 2}$, where $K_g$ is the $K$-value of the ground-state
band. In Fig.~\ref{fig5}, both these possible $\gamma$-bands with
$K=9/2$ and 1/2 are displayed. Since the two $\gamma$-bands
originate from the same intrinsic qp structure, the bandhead
energies of them are roughly the same, as seen in Fig.~\ref{fig5}.
As a rule, the one with the larger $K$ (here the $K=9/2$ one) is
energetically favored as the band starts to rotate at a higher spin
(here from $I=9/2$). $K=1/2$ $\gamma$-band lies at a higher
excitation energy for the entire spin regime as compared to the
favored $K=9/2$ band in Fig.~\ref{fig5}. The splitting between these
two bands is quite interesting and can provide a measure of the
interaction between qp and collective degrees of freedom. The
predicted $\gamma\gamma$-band lies between the $K=9/2$ and $K=1/2$
$\gamma$-bands and comes very close to the $K=9/2$ $\gamma$-band for
high-spin states.

The results of TPSM calculations after configuration mixing are
compared with the known experimental data in Fig.~\ref{fig6}, where
the yrast-, $\gamma$-, and $\gamma$$\gamma$-bands are shown. While
there exist experimental data for the yrast band for every isotope
up to quite high spin, the data for the $\gamma$- and
$\gamma\gamma$-bands are known only for $^{105}$Mo for which the
TPSM calculations are noted to reproduce the data reasonably well.
In particular, the bandheads of both $\gamma$- and
$\gamma\gamma$-bands are described correctly. The TPSM calculations
again predict two $\gamma$-bands with $K=1/2$ and 9/2 but we show
only the one with the larger K-value. The observed yrast bands for
all the studied odd-neutron systems clearly depict signature
splitting that increases with increasing angular momentum. This
splitting is well reproduced for all the nuclei in Fig.~\ref{fig6}
except for $^{109}$Ru where discrepancy is clearly noted, in
particular, for the low spin states. To have a better comparison
between the data and the TPSM calculations, yrast-, the level
energies of $\gamma$-, and $\gamma$$\gamma$-bands are compared in
Fig.~\ref{fig7} for $^{105}$MO. It is quite evident from this figure
that comparison between the TPSM calculated energies and the
experimental data is quite satisfactory.

To make the study complete, the calculated B(E2) values for the
studied odd-neutron isotopes are depicted in Table.~\ref{tab4}.
Presently, there is no data available for a comparison. The
predicted in-band transitions show a similar behavior for all the
studied nuclei with increasing trend as a function of
angular-momentum. The general trend is that Ru isotopes show
relatively smaller B(E2) values as compared to Mo, in particular,
for the high spin states and this corresponds to smaller deformation
values for Ru isotopes. It is also evident that for a given nucleus,
the $\gamma$- and $\gamma$$\gamma$-bands have slightly weaker
collectivity than the corresponding yrast one. Experimental
confirmation for these predictions are very much desired.

\section{Summary and Conclusions}

In the present work a systematic analysis of the band structures
observed in neutron-rich Mo- and Ru-isotopes has been performed
using the recently developed multi-qp TPSM approach. In this mass
region, rich band structures have been observed and this is one of
the few regions where $\gamma$- and $\gamma\gamma$-bands have been
observed up to quite high-spin. The advantage of the TPSM approach
is that it provides a unified microscopic description of the
collective and the qp excitation modes. It has been demonstrated
that, in general, TPSM results are in good agreement with the
available experimental data for energy levels and transitions. Among
them $^{105}$Mo is the only known odd-mass system where well
developed $\gamma$- and $\gamma\gamma$-bands have been observed and
it has been shown that TPSM approach provides a satisfactory
description of the known band structures for this isotope. However,
there appears to be a problem to reproduce the bandhead of the
$\gamma\gamma$-band for many nuclei studied in this work and this
problem was also noted in our earlier investigation for $^{103}$Nb
\cite{GJ10}. A possible reason for this discrepancy could be that
$\gamma\gamma$-band is having considerable mixing from the
vibrational degree of freedom for cases where large disagreement is
noted.  A possible way to include the vibrational mode is by
superimposing different $\gamma$-deformed solutions in the spirit of
projected generator coordinate method \cite{Ch13}. This may provide
a resolution to this problem and also may help improve the overall
description of the nuclear spectroscopic quantities using the TPSM
approach.
%In the present study, we have also evaluated the electromagnetic transitions along
%the yrast-bands. The available data on transitions is very scarce, except for $^{106}$Mo where
%the data on BE2 transitions is available up to I=12. It is noted that the observed drop in
%BE2 for this system is correctly reproduced by the theoretical calculations. For the studied odd-mass
%nuclei, there is no available data and the predicted transitions depict an increasing
%trend with angular-momentum.

%\begin{acknowledgments}
\section*{Acknowledgments}
The authors would like to acknowledge W. Nazarewicz for his
encouragement during the course of this work. Research at SJTU was
supported by the National Natural Science Foundation of China (Nos.
11575112, 11135005), by the 973 Program of China (No. 2013CB834401),
and by the Open Project Program of the State Key Laboratory of
Theoretical Physics, Institute of Theoretical Physics, Chinese
Academy of Sciences, China (No. Y5KF141CJ1).
%\end{acknowledgments}

\end{document}